\documentclass[reprint,superscriptaddress,aps,amsmath,amssymb]{revtex4-1}

\usepackage{graphicx}
\usepackage{placeins}
\usepackage{dcolumn}
\usepackage{bm}
\usepackage{color}
\usepackage{natbib}

\usepackage{graphicx}
\usepackage{hyperref}
\usepackage{amsmath}

\usepackage{titlesec} 
\titlespacing\section{0pt}{10pt}{4pt}
\titlespacing\subsection{0pt}{10pt}{2pt}
\titlespacing\subsubsection{0pt}{10pt}{1pt}

\begin{document}
\title{Rapid parameter estimation of discrete decaying signals using autoencoder networks}
\author{Jim C. Visschers} \email{jvisschers@uni-mainz.de}
\affiliation{Institut f\"ur Physik, Johannes Gutenberg Universit\"at-Mainz, 55128 Mainz, Germany}
\affiliation{Helmholtz-Institut Mainz, GSI Helmholtzzentrum f\"{u}r Schwerionenforschung GmbH, Mainz 55128, Germany}

\author{Dmitry Budker}
\affiliation{Institut f\"ur Physik, Johannes Gutenberg Universit\"at-Mainz, 55128 Mainz, Germany}
\affiliation{Helmholtz-Institut Mainz,  GSI Helmholtzzentrum f\"{u}r Schwerionenforschung GmbH, Mainz 55128, Germany}
\affiliation{Department of Physics, University of California, Berkeley, California 94720-300, USA}

\author{Lykourgos Bougas} \email{lybougas@uni-mainz.de}
\affiliation{Institut f\"ur Physik, Johannes Gutenberg Universit\"at-Mainz, 55128 Mainz, Germany}
\affiliation{Helmholtz-Institut Mainz, GSI Helmholtzzentrum f\"{u}r Schwerionenforschung GmbH, Mainz 55128, Germany}
\date{\today}



\begin{abstract}
In this work we demonstrate the use of neural networks for rapid extraction of signal parameters of discretely sampled signals. In particular, we use dense autoencoder networks to extract the parameters of interest from exponentially decaying signals and decaying oscillations. By using a three-stage training method and careful choice of the neural network size, we are able to retrieve the relevant signal parameters directly from the latent space of the autoencoder network at significantly improved rates compared to traditional algorithmic signal-analysis approaches. We show that the achievable precision and accuracy of this method of analysis is similar to conventional algorithm-based signal analysis  methods, by demonstrating that the extracted signal parameters are approaching their fundamental parameter estimation limit as provided by the Cram\'er-Rao bound. Furthermore, we demonstrate that autoencoder networks are able to achieve signal analysis, and, hence, parameter extraction, at rates of 75\,kHz, orders-of-magnitude faster than conventional techniques with similar precision. Finally, we explore the limitations of our approach, demonstrating that analysis rates of $>$200\,kHz are feasible with further optimization of the transfer rate between the data-acquisition system and data-analysis system.
\end{abstract}

\maketitle 
\tableofcontents
\newpage
\section{Introduction}
Machine learning (ML) is becoming a widespread method for the generation, and analysis of big data. ML uses networks of interconnected neurons (i.e., neural networks) that, much like real brains, recognize patterns in data structures \cite{bishop1995neural,hertz2018introduction}. Generally, these neural networks are first trained in situations where the desired action/output of the network is known and subsequently used in similar, real world situations. 
Research fields where a precise mathematical description of the problem is difficult -if not impossible- stand much to gain from implementing ML solutions. Areas such as image processing  \cite{gao2020flow,shamir2020intelligent} or text generation and analysis  \cite{brown2020language} have seen innovations that would not have been possible without the implementation of ML.

In physical sciences, ML-based techniques are becoming more and more widespread (see Ref. \cite{carleo2019machine} for an extended review). Besides unlocking completely new innovations, ML techniques have been utilized in situations where finding solutions to physical problems requires a lot of computing effort using conventional methods. For example, in fluid simulations where normally one is required to solve the Navier–Stokes equations, ML can be used to accurately predict the evolution of a fluid simulation \cite{sanchez2020learning}, while reducing the amount of computation time of these complex simulations significantly. Furthermore, neural networks have been used to simulate light scattering by multi-layer nanoparticles and the subsequent design of these nanoparticles using backpropagation \cite{peurifoy2018nanophotonic}. In spectroscopy, machine learning is used to accurately classify physical objects based on noisy/complex spectroscopic data \cite{wang2018classification,del2009duroc,gniadecka2004melanoma}. Another field where ML techniques find their way is in experiments with extremely high inference rates and low latency constraints, such as in high-performance detector triggers \cite{nottbeck2019implementation} (e.g. the ATLAS experiment at the LHC at CERN). In these experiments, field-programmable gate array (FPGA) implementations of ML techniques are used to create complex hardware triggers for events with sub-microsecond lifetimes. 

However, ML techniques are notoriously opaque, meaning that, although they deliver desired results, our understanding of how the results are obtained and how patterns in data are detected by the underlying neural network is very limited \cite{burrell2016machine}. Often, little or no information on the physical system that is solved/simulated can be gained by studying the neural network itself, which is a significant drawback of these techniques. Furthermore, basic questions regarding the minimal amount of training data, or the optimal size of the underlying neural network required to obtain good precision for a pre-defined problem remain unanswered \cite{carleo2019machine}.

Signal-parameter estimation is an integral part of both fundamental and applied research, with precision, accuracy and speed being crucial for the real-time observation and control of physical and chemical dynamic processes.
A multitude of research fields rely on determination of time constants and frequencies of decaying signals, examples of which include: nuclear magnetic resonance (NMR) \cite{gunther2013nmr} where molecular structures are analysed from precise determination of the frequency and decay time of the Larmor precession of nuclear spins in a magnetic field; free-induction-decay (FID) optical magnetometry \cite{savukov2005nmr,gemmel2010ultra,nikiel2014ultrasensitive,grujic2015sensitive,hunter2018free} where measurement sensitivities depend on the precision of the measurement of the precession of magnetized spins about a magnetic field; and cavity ring-down spectroscopy (CRDS) \cite{wheeler1998cavity,romanini1997cw,berden2000cavity,berden2009cavity,li2019simultaneous} that relies on the detection of variations of the photon lifetime in optical cavities to identify trace gasses, measure absorption cross-sections or observe chemical reactions in real time. Other cavity-enhanced methods, such as cavity ring-down polarimetry (CRDP) \cite{muller2000cavity,muller2002cavity,sofikitis2014evanescent,bougas2015chiral,dupre2015birefringence,visschers2020continuous,bougas2012cavity} and ellipsometry (CRDE) \cite{stamataki2013monitoring,sofikitis2013sensitivity,sofikitis2015microsecond}, measure birefringence and/or dichorism of an optical medium through the precise estimation of the signal-decay time and the polarization beat oscillations superimposed on such decaying signals.

All the aforementioned methods require some form of data processing to determine their relevant signal parameters. Optimizing sensitivity and minimizing computational costs is usually done by averaging multiple, consecutive measurements, sacrificing temporal resolution. Therefore, it is imperative that one considers the sampling and acquisition rates of an instrument and balances that against the computational costs (calculation time) to analyze the acquired data, the time scales of relevant dynamics to be observed, and the requirement to observe those dynamics in real time. In cavity-enhanced spectroscopy, CRDP/CRDE techniques in particular, the decay times of interest are typically in the $10^{-7}\,-\,10^{-5}$\,s range, a few orders of magnitude smaller than the typical decay times of NMR experiments ($10^{-2}-10^1$\,s) or FID optical magnetometry techniques ($10^{-3}-10^{1}$\,s). As such, analysis methods that are sufficiently fast for the real-time analysis of single measurements of NMR or FID magnetometry experiments do not have the capability to offer real-time analysis in situations where the relevant time scales are much smaller than 1\,ms.
Different time- and frequency-based computational methods for rapid parameter estimation have been demonstrated, and evaluated, for cavity ring-down (CRD) spectroscopy methods \cite{halmer2004fast,mazurenka2005fast, everest2008discrete}. Notably, Fourier-transform implementations on FPGAs have demonstrated data analysis rates as high as 4.4\,kHz \cite{bostrom2015discrete}. Several works discuss time- and frequency-domain analysis algorithms for damped sinusoidal signals \cite{aboutanios2009estimation,aboutanios2011estimating} and a recently published comparative study includes three analysis methods of discretely sampled damped sinusoidal signals in terms of their speed, and attainable accuracy and precision \cite{visschers2020rapid}.

In this work we demonstrate a ML based approach to extract the relevant signal parameters from experimentally relevant signals with well-defined functional form. More specifically, we use dense autoencoder networks to encode, extract parameters from, and subsequently reconstruct two types of discretely sampled, decaying signals: (1) exponentially decaying signals and (2) decaying oscillations. We evaluate the autoencoder network on its precision and accuracy in parameter extraction and compare its performance to the fundamental estimation limits of such signals given by the Cram\'{e}r-Rao lower bound (CRLB). We show that the dense autoencoder network is able to reach analysis rates of 75\,kHz of 1000 sample signals with cost-effective computational facilities. This makes the ML method ideal for implementations where computational capabilities come at a premium, such as fast, portable cavity-enhanced sensing instruments.
\section{Theory}
\label{Theory}
\begin{figure}
    \centering
    \includegraphics[width=0.7\linewidth]{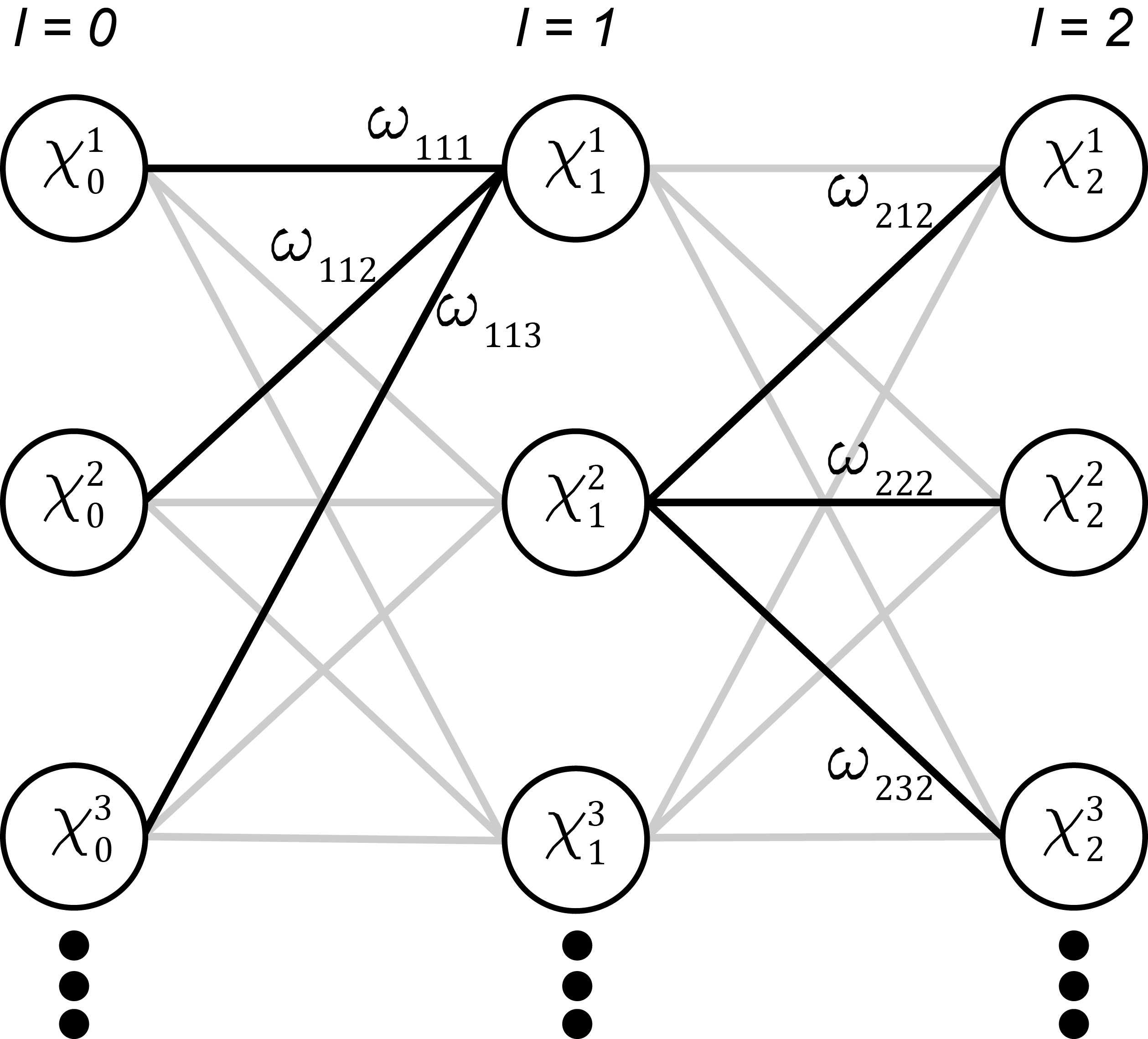}
    \caption{\small{Example of a dense neural network with $L = 2$ layers. The first layer, $l=0$ is the input layer, $l=1$ a hidden layer and $l=2$ is the output layer. $\chi^{x}_{y}$ specifies neuron $x$ in layer $y$. Each neuron contains a bias and layer specific activation function (not illustrated). $\omega_{y,x,z}$ denotes the weight (strength) between neuron $x$ in layer $y$ and neuron $z$ in layer $x-1$. For illustrative purposes, some connections have been highlighted and are shown with their corresponding weight. The output of the neuron depends on its inputs, weights and bias put through the activation function (Eq. \ref{Eq: NeuralNetwork feed forward}).}}
    \label{fig:Dense Neural network}
\end{figure}
\subsection{Dense autoencoder neural networks}
\label{sec. dense autoencoder NN}
A neural network is an ordered group of neurons that, much like the neurons in brains, communicate by signaling to each other. Each neuron has a number of incoming and outgoing connections from and to other neurons. Each connection has a strength (weight) and can be stimulating or inhibiting the response of the receiving neuron. To calculate the activity (value) of a neuron, one compares the sum of its weighted inputs to a neuron specific reference (bias). The difference is passed through an activation function which produces the output of the neuron \cite{hertz2018introduction,bishop1995neural}. Initially, the weights and biases within a neural network start off randomly. As the neural network is trained, these weights and biases are altered to optimize the performance of the network with respect to the task it is asked to perform. The manner in which neurons are ordered and connected influence the capabilities of the overall network and a variety of different network configurations with different purposes have been demonstrated \cite{NNZoo}.
A dense neural network is a network where the neurons are arranged in sequential layers and all the layers are fully connected. This means that every single neuron in one layer is connected to all neurons in the next layer. There are no neural connections within a layer or connections spanning multiple layers. An example of a dense neural network is shown in Fig.\,\ref{fig:Dense Neural network}. For such a neural network, one can calculate its output as:
\begin{equation}
    \begin{split}
    \text{for}\ l&: 0\longrightarrow L \\
    \vec{\chi}_{l+1} =&\, \mathcal{F}_l\left(\boldsymbol{\omega}_{l+1}\cdot\vec{\chi}_{l}-\vec{b}_{l}\right),
    \end{split}
    \label{Eq: NeuralNetwork feed forward}
\end{equation}
where $L$ is the number of layers in the neural network (excluding the input layer), $\vec{\chi}_{l}$ is the output of the network at layer $l$, $\boldsymbol{\omega}_{l}$ is the weights matrix, and $\vec{b}_{l}$ is the bias vector for the corresponding neurons of layer $l$. Finally, $\mathcal{F}_{l}$ is the activation function of the network layer that acts piece-wise on each neuron in layer $l$. The input layer ($l = 0$) has no bias or activation function. \\
\begin{figure}
    \centering
    \includegraphics[width=0.9\linewidth]{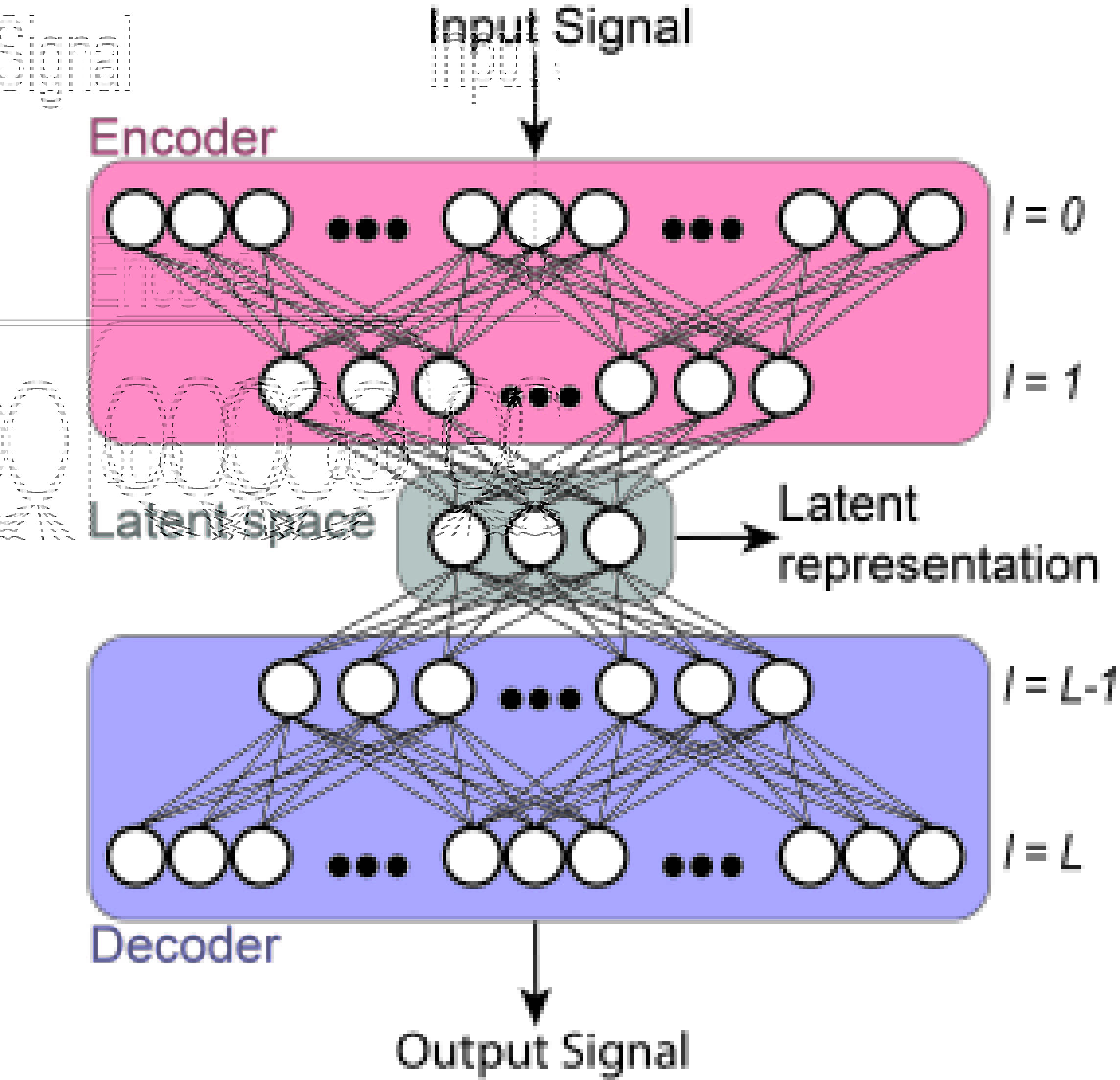}
    \caption{\small{Example of an autoencoder neural network with $L$ layers. Signals are fed to the encoder, which compresses the data into a limited number of parameters after which the decoder reconstructs the original data from the encoded parameters whilst minimizing the losses. The layer with the smallest number of neurons is called the latent space. From the latent space the encoded, or latent, representation of the input signal can be extracted.}}
    \label{fig:Autoencoder}
\end{figure}
Autoencoders are symmetric, hourglass-shaped neural networks  \cite{bourlard1988auto, demers1993non, hinton2006reducing, theis2017lossy,lu2013speech} (Fig.\,\ref{fig:Autoencoder}). This means that the number of neurons in the middle layer of the network is small compared to the number of neurons in the input and output layer of the neural network. The goal of the autoencoder network is to recreate its input signal. In the first half, the autoencoder network learns to encode the input data into a lower dimensional representation. In the second half, the autoencoder network learns to subsequently decode that lower dimensional representation and reconstruct the original signal with minimal losses. The middle layer of the network, in which the maximal compression of the signal takes place, is called the latent space. The latent representation of the original signal can be extracted from the latent space. Autoencoder networks are typically used in data compression \cite{theis2017lossy} and de-noising \cite{lu2013speech}. 

The choice of dimensionality of the latent space is ``\textit{problematic}'' \cite{demers1993non}. In order to achieve maximum data compression, the number of neurons in the latent space of the autoencoder network should be as small as possible. However one would also like to be able to reconstruct the data with minimal error. If the inherent dimensionality of the data is not known \textit{a priori} (as is typical) \cite{demers1993non,theis2017lossy,lu2013speech} but important, sparse, or variational autoencoder networks with Bayesian regularization methods should be used \cite{ng2011sparse}. Furthermore, inference from the latent space representation is nontrivial because the way a network ``learns'' to encode a signal in training is not necessarily unique.
Another way of structuring the latent space of an autoencoder network is by including the Kullback-Leibler (KL) divergence during training. However, KL divergence inevitably introduces a trade-off between the ability of the autoencoder network to reconstruct the original signal and a structured latent space \cite{Asperti2020Balancing}.

In this work, we examine how an autoencoder network consisting solely of dense layers can be employed for direct extraction of the  parameters of a signal with well-defined functional form from its latent signal-representation, as this becomes relevant in real-time signal analysis. To achieve such a task, we must make sure that (1) the dimensionality of the latent space matches the dimensionality of the input signal, and (2) that the autoencoder network encodes the input signal in a specific way where the latent representation of the signal matches the independent parameters of the signal. 

We are able to satisfy the first requirement because the number of parameters in our signal's functional form are known, which allows us to match the number of neurons in the latent space of the autoencoder network to the number of parameters of the function (Secs.\,\ref{Model Signals} \& \ref{Method Dense autoencoder networks}). Thereby we achieve efficient encoding without redundant neurons in the latent space, or redundant values in the latent representation of the signal. We are able to satisfy the second requirement by training the autoencoder network in a specific, three stage training method (Sec.\,\ref{Sec. Training Method}).

\subsection{Model signals}
\label{Model Signals}
We use autoencoder networks to estimate the parameters of two types of experimentally relevant model signals: (i) exponentially decaying signals and (ii) decaying oscillations.
\subsubsection{Exponentially decaying signals}
We start by investigating purely exponentially decaying signals. Such signals are encountered in a wide range of spectroscopic techniques, such as cavity ring down spectroscopy and cavity-enhanced sensing methods \cite{berden2000cavity,berden2009cavity,gagliardi2014cavity} and time-resolved fluorescence spectroscopy \cite{boens2007fluorescence,cundall2013time}.
An exponentially decaying signal can be characterized in terms of a model function as:
 \begin{equation}
    y(t) = A_0 \cdot e^{-t/\tau} + y_0(t),
    \label{EQ:ExpDecay}
\end{equation}
where $A_0$ is the initial amplitude and $\tau$ is the decay constant of the signal, $y_0(t)$ is the signal offset, and $t$ is the independent (time) variable that is discreetly sampled. Here, for simplicity and without loss of generality, we assume that $A_0=1$ and $\langle y_0(t)\rangle = 0$. Furthermore we restrict the investigation of noise contributions to the global offset parameter, i.e. $y_0(t)$. We model the noise to be normally distributed, i.e., $\langle y_{0}(t)\rangle = 0, \langle y_{0}^{2}(t)\rangle = \sigma_{y_{0}}^2$ and define the signal-to-noise ratio as: $\rm{SNR}=A_0/\sigma_{y_{0}}^2 = \sigma_{y_{0}}^{-2}$. Under realistic experimental conditions, signal-amplitude fluctuations can be incorporated into the SNR through $A_0$. This way, the dimensionality of the exponentially decaying signal is reduced to a single parameter that in most cases of interest carries the valuable information: the decay constant $\tau$.

\subsubsection{Decaying oscillations}
The second type of model signals we investigate are decaying oscillations, as these become relevant in a wide range of experimental techniques, particularly within nuclear magnetic resonance (NMR) \cite{gunther2013nmr}, optical magnetometry  \cite{savukov2005nmr,gemmel2010ultra,nikiel2014ultrasensitive,grujic2015sensitive,hunter2018free} and CRDP \cite{muller2000cavity,muller2002cavity,sofikitis2014evanescent,bougas2015chiral,dupre2015birefringence,spiliotis2020optical,visschers2020continuous,bougas2012cavity} and CRDE \cite{papadakis2011development,stamataki2013monitoring,sofikitis2013sensitivity,sofikitis2015microsecond}. An exponentially decaying oscillation can be characterized in terms of a model function as:
\begin{equation}
    y(t) = A_0 \cdot e^{-t/\tau} \cdot \text{cos}\left( 2\pi\cdot \text{f}\cdot t +\phi\right) + y_0(t),
    \label{Eq: FID}
\end{equation}
were t is the discretely sampled independent variable of the signal (time), $A_0$ is the oscillation's initial amplitude, $\tau$ is the decay constant of the signal's envelope, f is the frequency of the oscillation, and $\phi$ its phase. Finally $y_0(t)$ is the global signal offset. Again, we assume the signals amplitude to be normalized ($A_0 = 1$) and no global signal offset ($\langle y_0(t)\rangle = 0$), and SNR is defined as $\rm{SNR} = \sigma_{y_{0}}^{-2}$ identical to the exponentially decaying signals.  The experimentally relevant parameters $\tau$, f and $\phi$, are the free parameters we extract from the latent representation. We recognize that, similar to the case of pure exponential decays, the amplitude parameter $A_0$ could potentially be an experimentally relevant parameter for inspection purposes, and could be considered as a free parameter of the model signal.

\subsubsection{Cram\'er-Rao lower bound}
The fundamental limits for the statistical uncertainties of determining the decaying time-constant and oscillation frequency of pure and oscillating decaying signals are described by the Cram\'{e}r-Rao lower bound (CRLB). For the case of a decaying (oscillating) signal (Eq.\,\ref{EQ:ExpDecay} \& Eq.\,\ref{Eq: FID}) the CRLB \, \cite{yao1995cramer,gemmel2010ultra} sets the lower limit on the variance of both the decay constant estimator $\sigma_{\rm{\tau}}^2$ and the frequency estimator $\sigma_{\rm{f}}^2$. In general, the CRLB limit can be defined for any parameter extracted by an unbiased estimator. Here however, we focus on the two most important parameters for the experimental techniques of interest, e.g., \cite{berden2000cavity,berden2009cavity,gagliardi2014cavity,boens2007fluorescence,cundall2013time,nikiel2014ultrasensitive,hunter2018free,muller2000cavity,muller2002cavity,sofikitis2014evanescent,bougas2015chiral,dupre2015birefringence,spiliotis2020optical,visschers2020continuous,papadakis2011development,stamataki2013monitoring,sofikitis2013sensitivity,sofikitis2015microsecond, savukov2005nmr,gunther2013nmr,gemmel2010ultra,grujic2015sensitive, wheeler1998cavity,romanini1997cw,bougas2012cavity}.
    
    The relation between the variance limit of the decay-time estimator and the lower limit of the frequency estimator is given by: $\sigma_{\rm{\tau}}^2 = 2\pi \,\sigma_{\rm{f}}^2$. The CRLB for the frequency estimator $\sigma_{\rm{f}}^2$ is given by \cite{yao1995cramer,gemmel2010ultra}:
\begin{align}
    \sigma^2_{\rm{f}} = \dfrac{6}{(2\pi)^2\,\text{SNR}^2\, {\rm{f}}_{_\text{BW}} \, T_\text{m}^3}\, \xi\left(\tau/T_{\text{m}}\right),
    \label{EQ:CRLB}
\end{align}
where SNR is the signal-to-noise ratio of the decaying oscillating signal; f$_{_\text{BW}}$ is the sampling-rate-limited bandwidth of the measurement; $T_{\text{m}}$ is the measurement time window and, $\xi\left(\tau/T_{\text{m}}\right)$ is a correction factor that takes into account the signal decay, which is given by \cite{yao1995cramer,gemmel2010ultra}:
\begin{align}
    \xi(r) = \dfrac{\exp{\left(2/r\right)} - 1}{3r^3\,\text{cosh}\left(2/r\right)-3r\left(r^2+2\right)}.
\end{align}
The factor $\xi\left(\tau/T_{\text{m}}\right)$ serves as a compensation factor in Eq.\,\ref{EQ:CRLB} that penalizes measurement of the tails of the exponential decay when the signal has effectively died out. The validity of Eq.\,\ref{EQ:CRLB} hinges on the assumption that the period of the oscillation is shorter than the typical decay time ($\tau$) of the signal envelope, and that a sufficient number of oscillations occur in the measurement time window. Moreover, Eq.\,\ref{EQ:CRLB} dictates that any noise sources affecting the signal detection are contributing to the fundamental CRLB limit through their effect on the SNR of signal. 

In Ref.\, \cite{visschers2020continuous}, it was demonstrated that the CRLB limit is the appropriate estimator of the fundamental sensitivity of frequency-based measurements within the context of cavity-enhanced spectro-polarimetric techniques \cite{muller2000cavity,muller2002cavity,sofikitis2014evanescent,bougas2015chiral,dupre2015birefringence,spiliotis2020optical,visschers2020continuous,bougas2012cavity,papadakis2011development,stamataki2013monitoring,sofikitis2013sensitivity,sofikitis2015microsecond}.

\section{Methods}
\subsection{Simulated signals and data-sets}
We use simulated signals generated using the model functions presented in Eqs.\,\ref{EQ:ExpDecay}\,\&\,\ref{Eq: FID} to train and demonstrate the parameter-extraction capabilities of dense autoencoder networks.
For our simulated signals - both pure exponentially decaying and decaying oscillating ones -  we assume a total signal duration of 5\,$\mu$s and an effective bandwidth-limited sampling rate of 200\,MHz, resulting in discretely sampled signals with a length of 1000 samples. Such signals are frequently encountered in CRDP experiments \cite{visschers2020continuous}. 
We generate training data consisting out of multiple signals with varying experimentally relevant parameters. From the training data the autoencoder network learns the relationship between the shape of the signal and the values of the relevant parameters.
For exponentially decaying signals (Eq.\,\ref{EQ:ExpDecay}), the only parameter of interest is the decay constant $\tau$, which we vary between simulated signals as follows:
\begin{align}
\tau  = \left|\text{norm}\left(\mu_{\tau} = 1\,\mu s,\quad \zeta_{\tau} = 0.5\,\mu s\right)\right|,
\label{EQ: Tau Variation ExpDec}
\end{align}
where $\mu_\tau$ and $\zeta_\tau$ are the average and standard deviation of a normal distribution, from which we take the absolute value as a negative decay constant would lead to exponentially increasing signal. 
For the decaying oscillations (Eq.\,\ref{Eq: FID}) we vary the decay constant using the same process, but also vary the other two experimentally relevant parameters, the frequency and phase of the oscillation, f and $\phi$ respectively, using:
\begin{align}
\text{f} \,    &= \,\text{norm}\left(\mu_{\text{f}}\,\, = 3\,\text{MHz},\quad \zeta_{\text{f\,}} = 0.1\,\text{MHz}\right)   \label{Eq: FID f generation}\\
\phi  &= \,\text{norm}\left(\mu_{\phi} = 0,\quad\quad\quad\,\, \zeta_{\phi} = 0.1\right) \label{Eq: FID phi generation}
\end{align}
For the exponentially decaying signals a training data set consists of 200 signals, while for the decaying oscillations, a training data set consists of 1000 signals. These signals have a SNR$\,= 2^{20}$, to allow the autoencoder network to learn the features of each model signal and the signals variance under changing underlying parameters of interest without being hindered by noise artefacts.

\subsection{Dense autoencoder networks}
\label{Method Dense autoencoder networks}
The autoencoder networks that we use to reconstruct and extract the parameters from decaying signals, have an input layer and output layer of 1000 neurons, equal to the number of samples in a signal. However the number of neurons in the latent space of the autoencoder networks are different between autoencoder networks that analyze exponentially decaying signals and decaying oscillations. In the case of the exponentially decaying signals, we use a network with only a single neuron in the latent space, corresponding with the single parameter of interest that we wish to extract from the signal, i.e., $\tau$. In the case of the decaying oscillations, we need three neurons in the latent space of the network in order to be able to extract the three interesting parameters of that signal, i.e., $\tau, \textrm{f}, \phi$. 
Furthermore, the signal complexity of a decaying oscillation with three parameters is higher than a exponentially decaying signal with only one parameter of interest.
For this reason, to analyze decaying oscillations we increase the number of neurons in the layers of the encoder and decoder part of the autoencoder network (Fig.\,\ref{fig:Autoencoder}). 
The network we use to analyse the exponentially decaying signal has 5 total layers with 1000\,-\,50\,-\,1\,-\,50\,-\,1000 neurons per layer respectively. For the decaying oscillations, we use a network with 7 layers and 1000\,-\,50\,-\,10\,-\,3\,-\,10\,-\,50\,-\,1000 neurons per layer respectively. 
We choose to use a hyperbolic tangent as activation function for all relevant layers. This choice allows the neural network to express values between -1 and 1, which matches the maximal amplitude of the input and output signals. We create the aforementioned neural networks using the Tensorflow and Keras \cite{chollet2015keras} libraries in a  homemade Python script.

\subsection{Signal reconstruction and parameter extraction}
Autoencoder networks are used to reconstruct the original input data. However, our focus here is to use autoencoder networks for the rapid extraction of the signal parameters from exponentially decaying signals and decaying oscillations from the latent space of the autoencoder network. To do this we map the signal parameters of interest to a number between -1 and 1 that the autoencoder network can express in the latent space:
\begin{align}
    x_{\text{lat}} = \dfrac{ x - \mu_{x}}{3\times\zeta_{x}},
    \label{eq. param conversion}
\end{align}
where $x_{\text{lat}}$ is the latent representation of the value of parameter $x$ ($\tau$, f, and $\phi$). $\mu_{x}$ and $\zeta_{x}$ are the mean and standard deviation of the parameter variation defined in Eqs.\,\ref{EQ: Tau Variation ExpDec}-\ref{Eq: FID phi generation}. We use Eq.\,\ref{eq. param conversion} to create a desired latent representation of signal parameters during training, and to convert the latent representation back into the signal parameter when using the autoencoder network to analyse signals.
For a trained neural network, a signal only needs to be passed through the encoding part of the network, up to the latent space, in order to find its latent representation. To investigate the analysis speed of the neural networks, we only take into account the time it takes to encode the original signal into the latent representation.
\subsection{Training protocol}
The autoencoder networks are trained using a stochastic gradient descent (SGD) algorithm \cite{chollet2015keras} and following a three-stage training scheme. Even though the sizes of the autoencoder networks are different given the two different model signals, the training method for the autoencoder networks is the same:
\begin{enumerate}
    \item The complete autoencoder network is trained to recreate simulated input signals for 100 epochs.
    \item The encoder part of the autoencoder network, up to the latent space (Fig.\,\ref{fig:Autoencoder}), is trained to generate the desired latent-signal representation from simulated input signals. 
    \item Finally, the decoder part of the autoencoder network is trained with the desired latent-signal representation as input and the desired signals as output for 100 epochs.
\end{enumerate}
We repeat these steps 10 times after which a new training data set is generated to avoid over-fitting a single data set. In total, each network is trained on 10 different data sets. By implementing three distinct training steps we are able to train the neural network to not only recreate the original signal, but also to encode the signal in a specific way that allows extraction of the signal parameters from the latent space.

\begin{figure}
    \centering
    \includegraphics[width=1\linewidth]{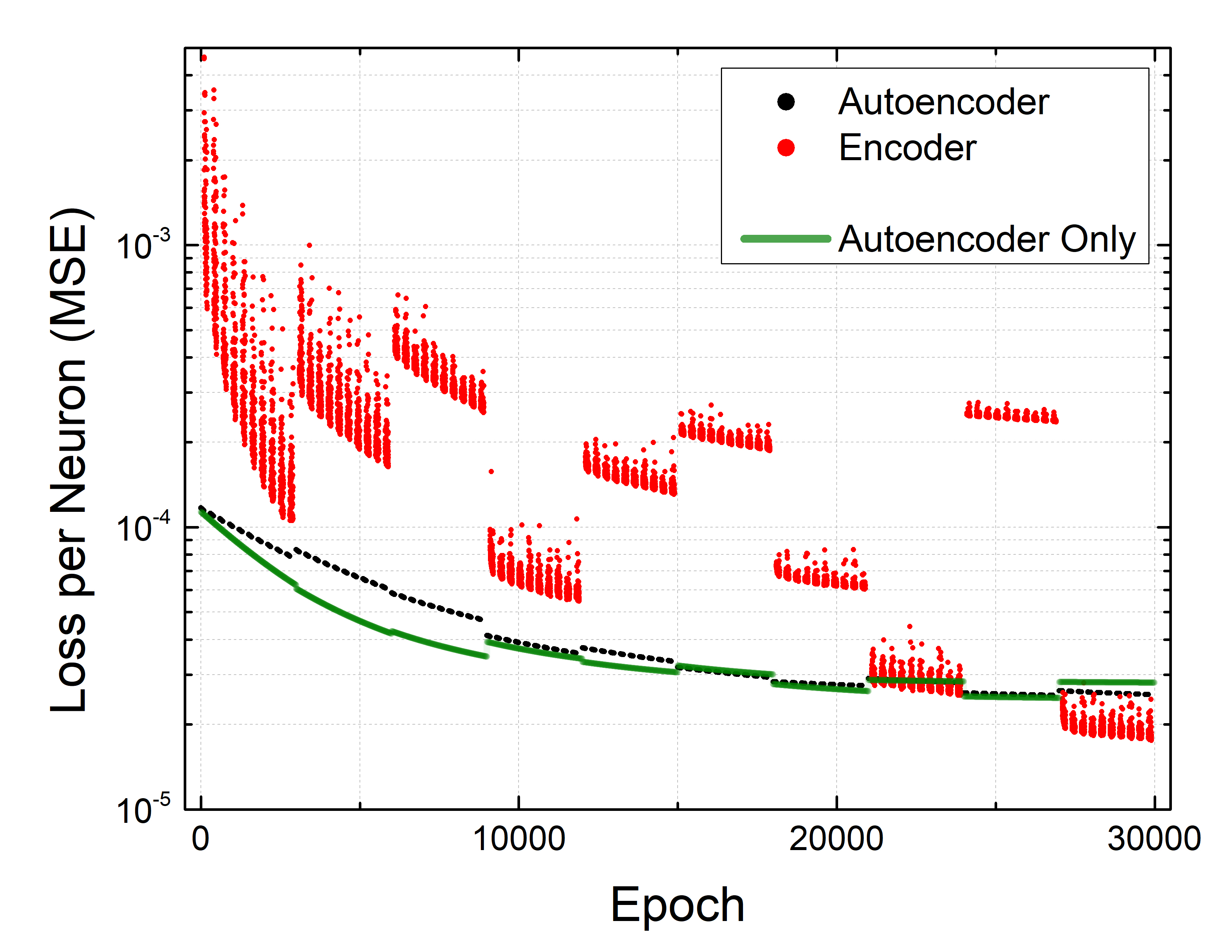}
    \caption{\small{Average mean squared error (MSE) loss per neuron as a function of training epochs for two networks. One network is trained using the three-stage training method outlined in Sec. \ref{Sec. Training Method} (dots) while the second network is trained in a typical fashion for an autoencoder network consisting of the first step only (dark-green line). The two networks have the same shape and size, are trained for an equal number of epochs using an equal amount of training data. The black and red dots correspond to the losses of the first network at training step 1 and 2 respectively (see text). The losses at training step 3 are omitted from the figure because they overlap with the losses of the first training step. The breaks in the losses of the encoder network training (red dots) when a new data set is generated are attributed to the small number of neurons in the latent space of the autoencoder network and the specifics of the small validation data set.}}
    \label{fig:3-stage training results}
\end{figure}

Figure \ref{fig:3-stage training results} compares the prediction error (loss) of two autoencoder networks during training. We measure the loss of each neuron in the output layer and the latent space using the mean squared error (MSE) \cite{chollet2015keras}. The smaller the loss of the autoencoder network, the better the network is in recreating the desired output.
The first network (red and black dots) is trained using our three-stage training method. The second network (dark-green line) is only trained on replicating the input data (first step). Both networks encode and reconstruct exponentially decaying signals described by Eq.\,\ref{EQ:ExpDecay}. We observe that both training methods converge to similar losses per neuron at the end of training for the complete autoencoder networks (black dots versus dark-green line), meaning that both networks, at the end of training, are able to reproduce the input data with a similar level of precision and accuracy. However, in the autoencoder network that is trained using our three-stage training method, we see a decrease in losses from the encoder part of the autoencoder network (red dots), indicating that by using the three-stage training method we can achieve a desired latent signal representation without sacrifice of the capability of the autoencoder network to reconstruct the original signal.
\label{Sec. Training Method}
\section{Results}
\subsection{Exponentially decaying signals}
\begin{figure*}[]
    \centering
    \includegraphics[width=1\linewidth]{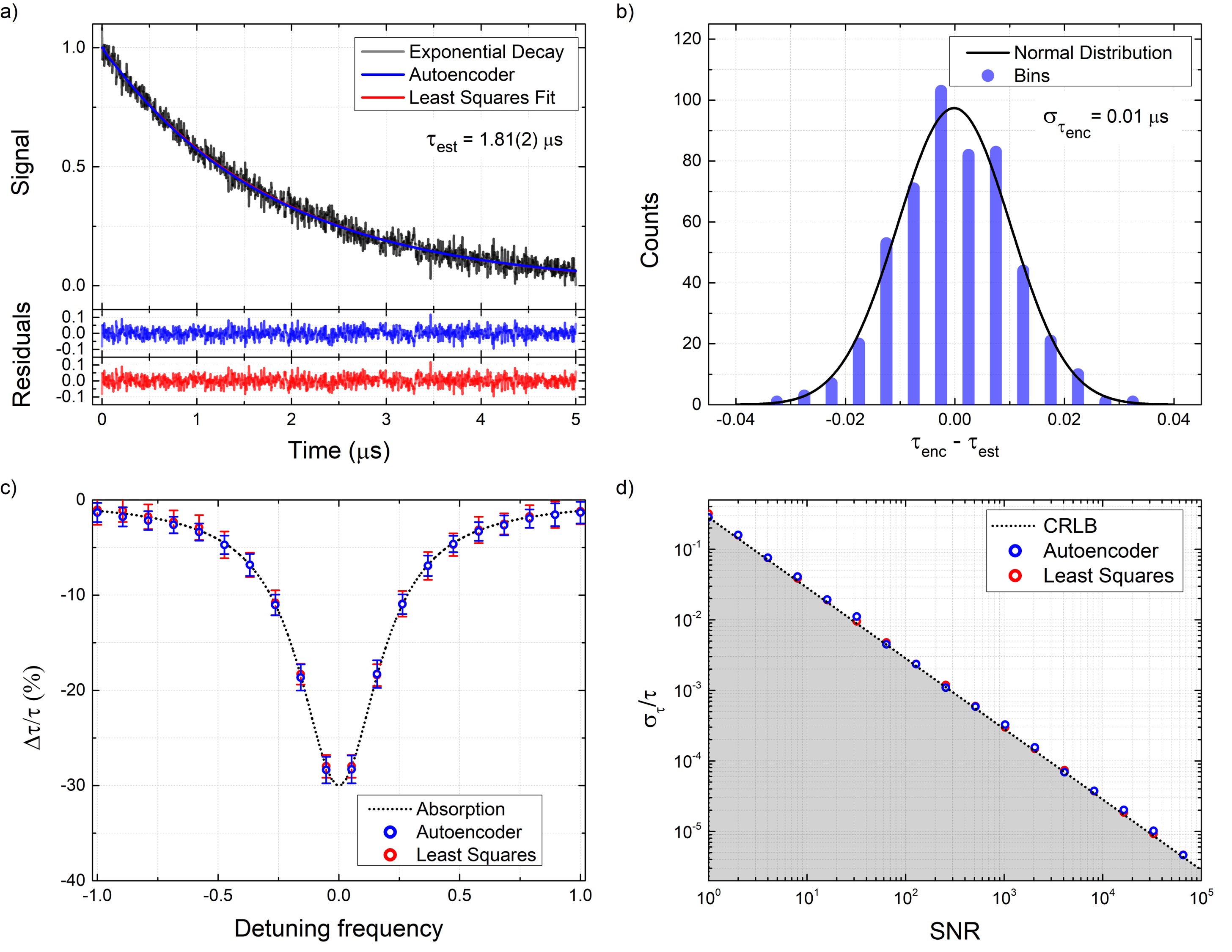}
    \caption{\small{Parameter extraction from an exponentially decaying signal using an autoencoder network. \textbf{(a)} Example of an exponentially decaying signal with 1000 samples and SNR$=2^5$ (black line) reconstructed by a trained autoencoder network (blue line) and analyzed using a least squares fitting method (red line). The residuals for each method are shown in the bottom insets of the figure with their corresponding colour. The decay constant estimated by the least squares method is $\tau_{\text{est}}\,=\,1.81(2)\,\mu$s. 
    \textbf{(b)} Histogram of 500 decay-constant estimates extracted from the latent space of the trained autoencoder network. Analysed signals have decay constants and SNR equal to the signal shown in (a). The value of the decay constant estimated by the least squares method, $\tau_{\text{est}}$, is subtracted from the decay constant extracted from the latent space of the autoencoder network, $\tau_{\text{enc}}$. The full width at half maximum (FWHM) of the fitted Gaussian distribution coincides with the uncertainty of the least squares method while the center of the distribution is not significantly different from 0. 
    \textbf{(c)} Accurate and precise signal parameter extraction using an autoencoder network is possible in the case of (relative) changes in the decay constant through a simulated absorption feature (dotted line) as one would expect in, for instance, cavity ring-down absorption spectroscopy when a laser is detuned from the resonance frequency of the absorption feature. The feature is  followed by both the trained autoencoder network (blue dots) and a least squares algorithm (red dots). The error bars represent the standard deviation of 100 parameter estimations from the latent space of the trained autoencoder network and the least squares algorithm respectively. 
    \textbf{(d)} Precision of a least squares algorithm and a trained autoencoder network on estimating the decay constant from the latent space over multiple orders of magnitude in SNR. The Cram\'er-Rao lower bound (CRLB) giving the fundamental estimation limit on the decay constant is also shown (dotted line) and the area below the fundamental estimation limit is greyed out.}}
    \label{fig:Results exponential decay}
\end{figure*}
Figure \ref{fig:Results exponential decay} shows the results for a trained autoencoder network reconstructing and extracting parameters from exponentially decaying signals. The ability of the autoencoder network to extract the signal parameters from the latent space is compared to the performance of an ordinary least squares algorithm \cite{visschers2020continuous}. In Fig.\,\ref{fig:Results exponential decay}(a) we show an example of a generated exponential decaying signal with SNR $=2^5$ and a randomly selected time constant ($\tau = 1.81\,\mu$s). We use a trained autoencoder network (Sec.\,\ref{Sec. Training Method}) to reconstruct this generated signal. In addition, we use a least-squares algorithm to analyze the generated signal, from which we extract the expected decay constant $\tau = 1.81(2)\,\mu$s, with a precision in accord to the CRLB limit (see related discussion in Ref. \cite{visschers2020rapid}). In Fig.\,\ref{fig:Results exponential decay}(b) we show a histogram of 500 estimates of the decay constant extracted from the latent space of the autoencoder network. By subtracting the least squares decay constant estimate ($\tau_{est}$) and fitting the histogram to a Gaussian distribution we show that there is no difference in accuracy between the two analysis methods as the center of the distribution is not significant from zero. The width of the distribution of the decay constant estimates coincides with the uncertainty of the least squares estimation method indicating that also the precision of the two methods is equal. In Fig.\,\ref{fig:Results exponential decay}(c) we demonstrate that the trained autoencoder network is able to, accurately and without loss in precision, follow signal changes. In particular, we simulate a spectral absorption feature having a characteristic dispersive (Lorentzian) profile. In this case, the decay constant changes by ~30\% and both the autoencoder network and the least squares algorithm are able to accurately detect such a change. . This situation is representative of a spectroscopy experiment investigating, for instance, absorption from gaseous species using CRDS  \cite{wahl2006ultra,li2019simultaneous}. In Fig.\,\ref{fig:Results exponential decay}(d) we show that the precision of the neural network is able to match the precision of the least squares algorithm and that both are limited by the CRLB over several orders of magnitude in SNR. It is crucial to emphasize that the network is able to accurately and precisely extract the correct signal parameters over a wide range of SNR values, despite the fact that we do not vary the SNR of the input training data during the training of the network.

\subsection{Decaying oscillations}
\begin{figure*}[ht]
    \centering
    \includegraphics[width=1\linewidth]{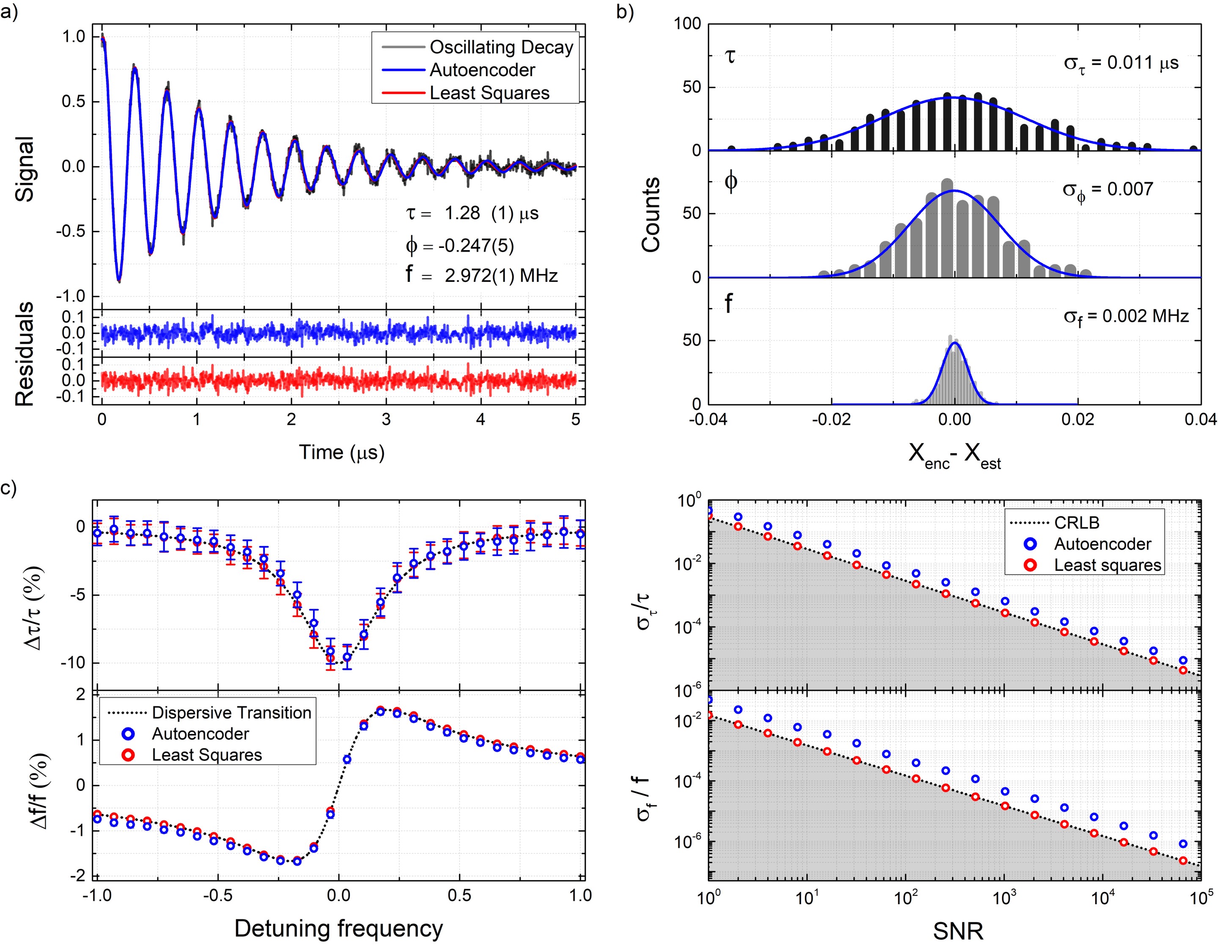}
    \caption{\small{Parameter extraction from decaying oscillating signals using an autoencoder network. \textbf{(a)} Example of a decaying oscillation (black line) with SNR\,$=2^5$ reconstructed by a trained autoencoder network (blue line) and analysed using a least squares fit (red line). The residuals for each method are shown in the bottom insets with their corresponding colour. The signal parameters found by the least squares fit are shown in the figure with their corresponding uncertainties.
    \textbf{(b)} Histograms of 500 signal parameter sets (top: $\tau$, middle: $\phi$ and bottom: f) extracted from the latent space of the autoencoder network. The analysed signals have the same parameters and SNR as the example signal shown in (a). For each of the signals parameters, we subtract the parameter estimate of the least squares fit X$_{est}$ from the parameter estimate extracted latent space of the autoencoder network X$_{enc}$. The width of the fitted Gaussian distributions coincide with the uncertainty of the least squares method for each individual parameter of the signal and the centers of the distributions are not significantly different form 0.
    \textbf{(c)} 
    Accurate and precise signal parameter extraction using an autoencoder network is possible in the case of (relative) changes in the decay constant through a simulated disperive absorption feature, or cotton effect, (dotted line) as one would expect in a cavity enhanced polarimetry experiment when a laser is detuned from the resonance frequency of the absorption feature. The feature is followed by both the trained autoencoder network (blue dots) and the least squares algorithm (red dots).The error bars represent the standard deviation of 100 parameter estimations from the latent space of the trained autoencoder network and the least squares algorithm respectively.
    \textbf{(d)} Precision of the trained autoencoder network (blue dots) and the least squares algorithm (red dots) in estimating both the frequency (f) and decay constant ($\tau$) over several orders of magnitude in SNR. The fundamental estimation limits of these parameters, given by the Cram\'{e}r-Rao lower bound (CRLB) is also shown (dotted line) and the area below the fundamental estimation limit is greyed out.}}
    \label{fig:Results oscillations}
\end{figure*}
In Fig.\,\ref{fig:Results oscillations} we present the results of a trained autoencoder network reconstructing and extracting parameters from decaying oscillations. In Fig.\,\ref{fig:Results oscillations}(a) we show an example of a decaying oscillating signal with SNR $=2^5$,  $\tau = 1.28\,\mu$s, $\phi = -0.243$, and f$=2.972$\,MHz. We fit the signal to Eq.\,\ref{Eq: FID} using a least squares algorithm and use a trained autoencoder (Sec.\,\ref{Sec. Training Method}) to reconstruct the signal. Figure \ref{fig:Results oscillations}(b) shows the histograms of 500 parameter estimates of $\tau$ , $\phi$ and f, extracted from the latent space of the autoencoder network. We extract the latent parameters $\tau_{\rm{lat}}$,\,$\phi_{\rm{lat}}$,\,$\rm{f}_{\rm{lat}}$ from the latent space of the autoencoder network and use Eq.\,\ref{eq. param conversion} to reconstruct the estimate of the signal parameters found by the autoencoder $\tau_{\rm{enc}}$,\,$\phi_{\rm{enc}}$,\,$\rm{f}_{\rm{enc}}$.
Following the same procedure described for the case of pure exponentially decaying signals, for each parameter we subtract the parameter's estimated value obtained through the least squares method and fit the histograms to a Gaussian distribution. Each distribution has a center that is not significant from zero and a width equal to the uncertainty of the fit parameters found by the least squares fit indicating that there is no difference in accuracy or precision between the two analysis methods.
In Fig.\,\ref{fig:Results oscillations}(c) we demonstrate that the trained autoencoder network is able to, accurately and without loss in precision, follow signal changes. In particular, in Fig.\,\ref{fig:Results oscillations}(c) we simulate a spectral feature having a characteristic change in both its absorption and dispersion (e.g., Cotton effect). This results in signals whose frequency and decay constant change in a correlated way. Finally, in Fig.\,\ref{fig:Results oscillations}(d) we show that the precision of the parameter estimates from the latent space of the trained autoencoder network follows the precision of the least squares method over five orders of magnitude in SNR. Moreover, the precision of the parameters extracted from the latent space of the autoencoder network approaches the fundamental estimation limit as given by the CRLB. 

\subsection{Complexity vs. calculation time}
\begin{figure}[ht!]
    \centering
    \includegraphics[width=1\linewidth]{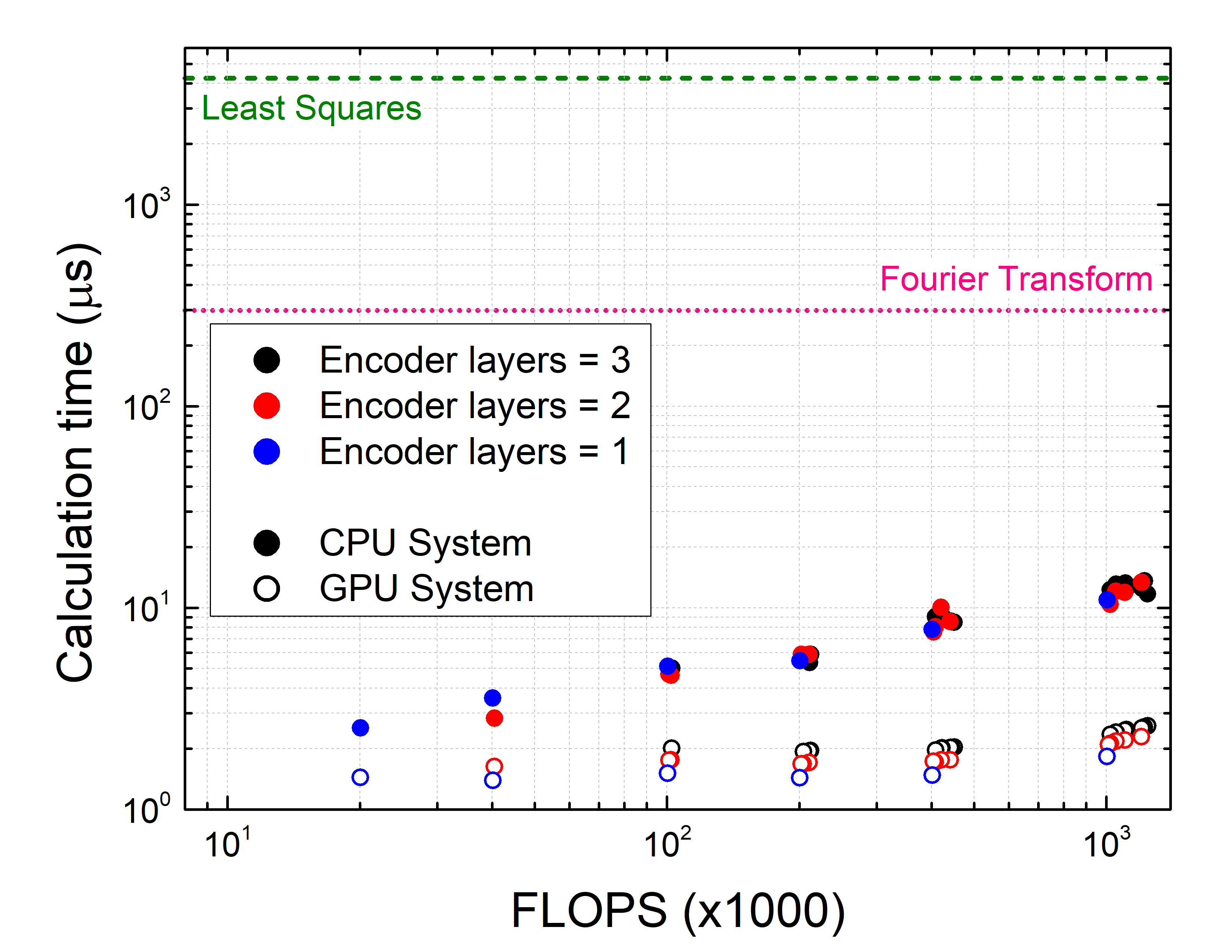}
    \caption{\small{Calculation time required by the encoder part of the autoencoder network to extract the latent parameters of a single damped oscillating signal as a function of floating point operations (FLOPS) required to calculate the network using a CPU- (solid dots) and a GPU- (open dots) based system. The analysed signals consist of 1000 samples. The networks are varied in size, and thereby the required number of FLOPS needed to calculate latent signal representation. We vary the size of the network by varying the number of layers in the encoder network and the number of neurons within each layer. The lengths of these layers range from 10 to 500 neurons, where the smallest network investigated has a single layer with 10 neurons, and the largest network has 3 layers with 500, 200 and 100 neurons between the input layer and the latent space respectively. The dashed and dotted horizontal lines represent the calculation times required to analyse the same signal using a least squares algorithm and non-iterative Fast Fourier transform method, respectively \cite{visschers2020rapid}.}}
    \label{fig:Results speed}
\end{figure}
Analysis speed is crucial for the real-time investigation and control of fast processes. Faster methods of analysis allow for large data stream to be analysed quicker, driving down computational costs. 

For our particular cases of interest, we recently showed (Ref. \cite{visschers2020rapid}) that non-iterative FFT methods require $\sim 300\,\mu$s of calculation time whereas a least-squares fitting algorithms requires $>$ 1\,ms of calculation time under identical signal conditions as we present in this work. Similarly, other works have shown that FPGA- based  systems running FFT algorithms are able to determine the decay constant of exponentially decaying signals with analysis rates of 4.4\,kHz \cite{bostrom2015discrete}. Here, we show that trained autoencoder networks are able to analyse both exponentially decaying signals and decaying oscillations significantly faster than any previously reported method. We compare the calculation speed of the encoder part of trained autoencoder networks of different sizes on different systems: a central processing unit (CPU-) and a graphics-processing unit (GPU-) based system. The CPU-based system is based around an Intel(R) Xeon(R) W-2123, 3.6\,GHz processing unit with access to 16\,GB of random-access memory (RAM) with a frequency of 1330 MHz, while the GPU system is based on a NVIDIA Quadro p5000 graphics board.

In Fig.\,\ref{fig:Results speed} we present the calculation time required by the encoder part of an autoencoder network to analyse a single decaying oscillation signal for varying sizes of networks. For our estimation of the overall CPU calculation time (Fig.\,\ref{fig:Results speed}) we count the matrix multiplications, the adding/subtracting of the network biases and the application of the activation function. Other prepossessing operations for the CPU-based system, such as data collection, take between 3.6 - 4.9\,$\mu$s per signal and are not included. For our estimation of the overall GPU calculation time we include the aforementioned operations (such as matrix multiplications, the adding/subtracting of the network biases, and the application of the activation function) and the transfer time of the results of the encoder onto the RAM of the computer. We do not include, however, the time it takes for the original signal to be transferred from the RAM onto the GPU memory ($\sim5\,\mu$s) or any other pre/post-processing operations conducted by the CPU of the system ($\sim6\,\mu$s). 
For the GPU-based system there is a clear difference in calculation time between neural networks with different numbers of layers in the encoder network, something we do not observe for the same neural networks on the CPU-based system.  The results we present in Figs.\,\ref{fig:Results exponential decay}\,-\,\ref{fig:Results oscillations} are obtained with autoencoder networks that require the calculation of $\sim$100,000 floating point operations (FLOPS) to reach the latent space representation of the signal. Using the trained autoencoder networks we are able to achieve analysis rates upwards of 75\,kHz using the GPU-based system. This includes the data transfer time, the pre/post processing operations conducted by the CPU, and the actual calculation time required of the autoencoder network itself. By optimizing the transfer rate between the data-acquisition system and data-analysis system and minimizing the prepossessing operations, down to a combined $\sim3\,\mu$s, the method of analysis we propose in this paper has the capabilities to reach analysis rates of $>$200\,kHz, in which we include the $<$2$\,\mu$s calculation time required by the system.

\section{Outlook and conclusion}
In summary we have shown that we can accurately and precisely extract the signal parameters of decaying signals using simple autoencoder networks. We demonstrate that our approach is orders of magnitude faster than conventional algorithmic methods (e.g., least-squares or FFT), regardless of CPU- or GPU implementation of the neural network. We demonstrate analysis rates upwards of 75\,kHz for signals with 1000 samples, and illustrate that analysis rates of $>$200$\,$KHz are feasible with optimization of data transfer speed between a data-acquisition and data-analysis device, which would allow for real time signal analysis rates of $>$200\,kHz using state-of-the-art GSa/s sampling rates. Such capabilities can enable the real-time signal analysis of, for instance, CRDP signals at $>200\,$KHz rates using state-of-the-art acquisition modules that have GSa/s sampling rates.

Concluding, we wish to note that the methodology of signal parameter extraction directly from the latent space of dense autoencoder networks could be applicable to other signal types that currently use fitting models for parameter extraction. Presently, neural networks are used for the classification of spectroscopic data \cite{wang2018classification,del2009duroc,gniadecka2004melanoma}, however, our approach can be directly implemented to analyse spectroscopic data for quantitative rather than qualitative results. 
Signals of higher complexity, such as signals with additional decay constants or frequency components, will require larger networks and larger training data sets for the network. If the number of extra parameters is known, a similar technique to what we demonstrate can be employed. However, if the number of additional parameters is unknown, a  regularization method should be used to adequately choose the number of latent space parameters of the network to analyse the signal.

\section*{Acknowledgments}
JCV and LB are grateful to Christian Schmitt and Michael Everest for their help and support and specially thank Sharon van Etten for insightful discussions. This work was supported by the European Commission Horizon 2020, project ULTRACHIRAL (Grant No. FETOPEN-737071).

\bibliography{main}

\begin{thebibliography}{57}
\providecommand{\natexlab}[1]{#1}
\providecommand{\url}[1]{{#1}}
\providecommand{\urlprefix}{}
\expandafter\ifx\csname urlstyle\endcsname\relax
  \providecommand{\doi}[1]{DOI~\discretionary{}{}{}#1}\else
  \providecommand{\doi}{DOI~\discretionary{}{}{}\begingroup
  \urlstyle{rm}\Url}\fi
\providecommand{\eprint}[2][]{\url{#2}}

\bibitem[{Bishop et~al.(1995)}]{bishop1995neural}
Bishop CM, et~al. (1995) Neural networks for pattern recognition. Oxford
  university press

\bibitem[{Hertz(2018)}]{hertz2018introduction}
Hertz JA (2018) Introduction to the theory of neural computation. CRC Press

\bibitem[{Gao et~al.(2020)Gao, Saraf, Huang, and Kopf}]{gao2020flow}
Gao C, Saraf A, Huang JB, Kopf J (2020) Flow-edge guided video completion. In:
  European Conference on Computer Vision, Springer, pp 713--729,
  \urlprefix\url{https://arxiv.org/abs/2009.01835}

\bibitem[{Shamir et~al.(2020)Shamir, Mitra, Umetani, and
  Koyama}]{shamir2020intelligent}
Shamir A, Mitra NJ, Umetani N, Koyama Y (2020) Intelligent tools for creative
  graphics. In: ACM SIGGRAPH 2020 Courses, pp 1--11,
  \urlprefix\url{https://doi.org/10.1145/3388769.3407498}

\bibitem[{Brown et~al.(2020)Brown, Mann, Ryder, Subbiah, Kaplan, Dhariwal,
  Neelakantan, Shyam, Sastry, Askell et~al.}]{brown2020language}
Brown TB, Mann B, Ryder N, Subbiah M, Kaplan J, Dhariwal P, Neelakantan A,
  Shyam P, Sastry G, Askell A, et~al. (2020) Language models are few-shot
  learners. arXiv preprint arXiv:200514165
  \urlprefix\url{https://arxiv.org/abs/2005.14165}

\bibitem[{Carleo et~al.(2019)Carleo, Cirac, Cranmer, Daudet, Schuld, Tishby,
  Vogt-Maranto, and Zdeborov{\'a}}]{carleo2019machine}
Carleo G, Cirac I, Cranmer K, Daudet L, Schuld M, Tishby N, Vogt-Maranto L,
  Zdeborov{\'a} L (2019) Machine learning and the physical sciences. Reviews of
  Modern Physics 91(4):045002,
  \urlprefix\url{https://doi.org/10.1103/RevModPhys.91.045002}

\bibitem[{Sanchez-Gonzalez et~al.(2020)Sanchez-Gonzalez, Godwin, Pfaff, Ying,
  Leskovec, and Battaglia}]{sanchez2020learning}
Sanchez-Gonzalez A, Godwin J, Pfaff T, Ying R, Leskovec J, Battaglia P (2020)
  Learning to simulate complex physics with graph networks. In: International
  Conference on Machine Learning, PMLR, pp 8459--8468,
  \urlprefix\url{http://proceedings.mlr.press/v119/sanchez-gonzalez20a.html}

\bibitem[{Peurifoy et~al.(2018)Peurifoy, Shen, Jing, Yang, Cano-Renteria,
  DeLacy, Joannopoulos, Tegmark, and
  Solja{\v{c}}i{\'c}}]{peurifoy2018nanophotonic}
Peurifoy J, Shen Y, Jing L, Yang Y, Cano-Renteria F, DeLacy BG, Joannopoulos
  JD, Tegmark M, Solja{\v{c}}i{\'c} M (2018) Nanophotonic particle simulation
  and inverse design using artificial neural networks. Science advances
  4(6):eaar4206, \urlprefix\url{https://doi.org/10.1126/sciadv.aar4206}

\bibitem[{Wang et~al.(2018)Wang, Liao, Zheng, Xue, and
  Peng}]{wang2018classification}
Wang J, Liao X, Zheng P, Xue S, Peng R (2018) Classification of chinese herbal
  medicine by laser-induced breakdown spectroscopy with principal component
  analysis and artificial neural network. Analytical letters 51(4):575--586,
  \urlprefix\url{https://doi.org/10.1080/00032719.2017.1340949}

\bibitem[{Del~Moral et~al.(2009)Del~Moral, Guill{\'e}n, Del~Moral, O’valle,
  Mart{\'\i}nez, and Del~Moral}]{del2009duroc}
Del~Moral F, Guill{\'e}n A, Del~Moral L, O’valle F, Mart{\'\i}nez L,
  Del~Moral R (2009) Duroc and iberian pork neural network classification by
  visible and near infrared reflectance spectroscopy. Journal of Food
  Engineering 90(4):540--547,
  \urlprefix\url{https://doi.org/10.1016/j.jfoodeng.2008.07.027}

\bibitem[{Gniadecka et~al.(2004)Gniadecka, Philipsen, Wessel, Gniadecki, Wulf,
  Sigurdsson, Nielsen, Christensen, Hercogova, Rossen
  et~al.}]{gniadecka2004melanoma}
Gniadecka M, Philipsen PA, Wessel S, Gniadecki R, Wulf HC, Sigurdsson S,
  Nielsen OF, Christensen DH, Hercogova J, Rossen K, et~al. (2004) Melanoma
  diagnosis by raman spectroscopy and neural networks: structure alterations in
  proteins and lipids in intact cancer tissue. Journal of investigative
  dermatology 122(2):443--449,
  \urlprefix\url{https://doi.org/10.1046/j.0022-202X.2004.22208.x}

\bibitem[{Nottbeck et~al.(2019)Nottbeck, Schmitt, and
  B{\"u}scher}]{nottbeck2019implementation}
Nottbeck N, Schmitt C, B{\"u}scher V (2019) Implementation of high-performance,
  sub-microsecond deep neural networks on {FPGA}s for trigger applications.
  Journal of Instrumentation 14(09):P09014,
  \urlprefix\url{https://doi.org/10.1088/1748-0221/14/09/P09014}

\bibitem[{Burrell(2016)}]{burrell2016machine}
Burrell J (2016) How the machine ‘thinks’: Understanding opacity in machine
  learning algorithms. Big Data \& Society 3(1):2053951715622512,
  \doi{https://doi.org/10.1177/2053951715622512}

\bibitem[{G{\"u}nther(2013)}]{gunther2013nmr}
G{\"u}nther H (2013) {NMR} spectroscopy: basic principles, concepts and
  applications in chemistry. John Wiley \& Sons

\bibitem[{Savukov and Romalis(2005)}]{savukov2005nmr}
Savukov I, Romalis M (2005) {NMR} detection with an atomic magnetometer.
  Physical review letters 94(12):123001,
  \urlprefix\url{https://doi.org/10.1103/PhysRevLett.94.123001}

\bibitem[{Gemmel et~al.(2010)Gemmel, Heil, Karpuk, Lenz, Ludwig, Sobolev,
  Tullney, Burghoff, Kilian, Knappe-Gr{\"u}neberg et~al.}]{gemmel2010ultra}
Gemmel C, Heil W, Karpuk S, Lenz K, Ludwig C, Sobolev Y, Tullney K, Burghoff M,
  Kilian W, Knappe-Gr{\"u}neberg S, et~al. (2010) Ultra-sensitive magnetometry
  based on free precession of nuclear spins. The European Physical Journal D
  57(3):303--320, \urlprefix\url{https://doi.org/10.1140/epjd/e2010-00044-5}

\bibitem[{Nikiel et~al.(2014)Nikiel, Bl{\"u}mler, Heil, Hehn, Karpuk, Maul,
  Otten, Schreiber, and Terekhov}]{nikiel2014ultrasensitive}
Nikiel A, Bl{\"u}mler P, Heil W, Hehn M, Karpuk S, Maul A, Otten E, Schreiber
  LM, Terekhov M (2014) Ultrasensitive $^{3}${H}e magnetometer for measurements
  of high magnetic fields. The European Physical Journal D 68(11):1--12,
  \urlprefix\url{https://doi.org/10.1140/epjd/e2014-50401-3}

\bibitem[{Gruji{\'c} et~al.(2015)Gruji{\'c}, Koss, Bison, and
  Weis}]{grujic2015sensitive}
Gruji{\'c} ZD, Koss PA, Bison G, Weis A (2015) A sensitive and accurate atomic
  magnetometer based on free spin precession. The European Physical Journal D
  69(5):135, \urlprefix\url{https://doi.org/10.1140/epjd/e2015-50875-3}

\bibitem[{Hunter et~al.(2018)Hunter, Piccolomo, Pritchard, Brockie, Dyer, and
  Riis}]{hunter2018free}
Hunter D, Piccolomo S, Pritchard J, Brockie N, Dyer T, Riis E (2018)
  Free-induction-decay magnetometer based on a microfabricated cs vapor cell.
  Physical Review Applied 10(1):014002,
  \urlprefix\url{https://doi.org/10.1103/PhysRevApplied.10.014002}

\bibitem[{Wheeler et~al.(1998)Wheeler, Newman, Orr-Ewing, and
  Ashfold}]{wheeler1998cavity}
Wheeler MD, Newman SM, Orr-Ewing AJ, Ashfold MN (1998) Cavity ring-down
  spectroscopy. Journal of the Chemical Society, Faraday Transactions
  94(3):337--351, \urlprefix\url{https://doi.org/10.1039/A707686J}

\bibitem[{Romanini et~al.(1997)Romanini, Kachanov, Sadeghi, and
  Stoeckel}]{romanini1997cw}
Romanini D, Kachanov A, Sadeghi N, Stoeckel F (1997) {CW} cavity ring down
  spectroscopy. Chemical Physics Letters 264(3-4):316--322,
  \urlprefix\url{https://doi.org/10.1016/S0009-2614(96)01351-6}

\bibitem[{Berden et~al.(2000)Berden, Peeters, and Meijer}]{berden2000cavity}
Berden G, Peeters R, Meijer G (2000) Cavity ring-down spectroscopy:
  Experimental schemes and applications. International reviews in physical
  chemistry 19(4):565--607

\bibitem[{Berden and Engeln(2009)}]{berden2009cavity}
Berden G, Engeln R (2009) Cavity ring-down spectroscopy: techniques and
  applications. John Wiley \& Sons,
  \urlprefix\url{https://doi.org/10.1080/014423500750040627}

\bibitem[{Li et~al.(2019)Li, Hu, Xie, Chen, Liu, Liang, Wang, Wang, Wang, Lin
  et~al.}]{li2019simultaneous}
Li Z, Hu R, Xie P, Chen H, Liu X, Liang S, Wang D, Wang F, Wang Y, Lin C,
  et~al. (2019) Simultaneous measurement of {NO} and {NO}$_{2}$ by a
  dual-channel cavity ring-down spectroscopy technique. Atmospheric Measurement
  Techniques 12(6):3223--3236,
  \urlprefix\url{https://doi.org/10.5194/amt-12-3223-2019}

\bibitem[{M{\"u}ller et~al.(2000)M{\"u}ller, Wiberg, and
  Vaccaro}]{muller2000cavity}
M{\"u}ller T, Wiberg KB, Vaccaro PH (2000) Cavity ring-down polarimetry
  ({CRDP}): a new scheme for probing circular birefringence and circular
  dichroism in the gas phase. The Journal of Physical Chemistry A
  104(25):5959--5968, \urlprefix\url{https://doi.org/10.1021/jp000705n}

\bibitem[{M{\"u}ller et~al.(2002)M{\"u}ller, Wiberg, Vaccaro, Cheeseman, and
  Frisch}]{muller2002cavity}
M{\"u}ller T, Wiberg KB, Vaccaro PH, Cheeseman JR, Frisch MJ (2002) Cavity
  ring-down polarimetry ({CRDP}): theoretical and experimental
  characterization. JOSA B 19(1):125--141,
  \urlprefix\url{https://doi.org/10.1364/JOSAB.19.000125}

\bibitem[{Sofikitis et~al.(2014)Sofikitis, Bougas, Katsoprinakis, Spiliotis,
  Loppinet, and Rakitzis}]{sofikitis2014evanescent}
Sofikitis D, Bougas L, Katsoprinakis GE, Spiliotis AK, Loppinet B, Rakitzis TP
  (2014) Evanescent-wave and ambient chiral sensing by signal-reversing cavity
  ringdown polarimetry. Nature 514(7520):76--79,
  \urlprefix\url{https://doi.org/10.1038/nature13680}

\bibitem[{Bougas et~al.(2015)Bougas, Sofikitis, Katsoprinakis, Spiliotis,
  Tzallas, Loppinet, and Rakitzis}]{bougas2015chiral}
Bougas L, Sofikitis D, Katsoprinakis GE, Spiliotis AK, Tzallas P, Loppinet B,
  Rakitzis TP (2015) Chiral cavity ring down polarimetry: Chirality and
  magnetometry measurements using signal reversals. The Journal of Chemical
  Physics 143(10):09B603\_1, \urlprefix\url{https://doi.org/10.1063/1.4930109}

\bibitem[{Dupr{\'e}(2015)}]{dupre2015birefringence}
Dupr{\'e} P (2015) Birefringence-induced frequency beating in high-finesse
  cavities by continuous-wave cavity ring-down spectroscopy. Physical Review A
  92(5):053817, \urlprefix\url{https://doi.org/10.1103/PhysRevA.92.053817}

\bibitem[{Visschers et~al.(2020)Visschers, Tretiak, Budker, and
  Bougas}]{visschers2020continuous}
Visschers JC, Tretiak O, Budker D, Bougas L (2020) Continuous-wave cavity
  ring-down polarimetry. The Journal of Chemical Physics 152(16):164202,
  \urlprefix\url{https://doi.org/10.1063/5.0004476}

\bibitem[{Bougas et~al.(2012)Bougas, Katsoprinakis, Von~Klitzing, Sapirstein,
  and Rakitzis}]{bougas2012cavity}
Bougas L, Katsoprinakis G, Von~Klitzing W, Sapirstein J, Rakitzis T (2012)
  Cavity-enhanced parity-nonconserving optical rotation in metastable xe and
  hg. Physical review letters 108(21):210801,
  \urlprefix\url{https://doi.org/10.1103/PhysRevLett.108.210801}

\bibitem[{Stamataki et~al.(2013)Stamataki, Papadakis, Everest, Tzortzakis,
  Loppinet, and Rakitzis}]{stamataki2013monitoring}
Stamataki K, Papadakis V, Everest MA, Tzortzakis S, Loppinet B, Rakitzis TP
  (2013) Monitoring adsorption and sedimentation using evanescent-wave cavity
  ringdown ellipsometry. Applied Optics 52(5):1086--1093

\bibitem[{Sofikitis et~al.(2013)Sofikitis, Stamataki, Everest, Papadakis,
  Stehle, Loppinet, and Rakitzis}]{sofikitis2013sensitivity}
Sofikitis D, Stamataki K, Everest MA, Papadakis V, Stehle JL, Loppinet B,
  Rakitzis TP (2013) Sensitivity enhancement for evanescent-wave sensing using
  cavity-ring-down ellipsometry. Optics Letters 38(8):1224--1226,
  \urlprefix\url{https://doi.org/10.1364/OL.38.001224}

\bibitem[{Sofikitis et~al.(2015)Sofikitis, Spiliotis, Stamataki, Katsoprinakis,
  Bougas, Samartzis, Loppinet, Rakitzis, Surligas, and
  Papadakis}]{sofikitis2015microsecond}
Sofikitis D, Spiliotis A, Stamataki K, Katsoprinakis G, Bougas L, Samartzis P,
  Loppinet B, Rakitzis T, Surligas M, Papadakis S (2015) Microsecond-resolved
  sdr-based cavity ring down ellipsometry. Applied Optics 54(18):5861--5865,
  \urlprefix\url{https://doi.org/10.1364/AO.54.005861}

\bibitem[{Halmer et~al.(2004)Halmer, von Basum, Hering, and
  M{\"u}rtz}]{halmer2004fast}
Halmer D, von Basum G, Hering P, M{\"u}rtz M (2004) Fast exponential fitting
  algorithm for real-time instrumental use. Review of scientific instruments
  75(6):2187--2191, \urlprefix\url{https://doi.org/10.1063/1.1711189}

\bibitem[{Mazurenka et~al.(2005)Mazurenka, Wada, Shillings, Butler, Beames, and
  Orr-Ewing}]{mazurenka2005fast}
Mazurenka M, Wada R, Shillings A, Butler T, Beames J, Orr-Ewing A (2005) Fast
  fourier transform analysis in cavity ring-down spectroscopy: application to
  an optical detector for atmospheric {NO}$_{2}$. Applied\ Physics\ B
  81(1):135--141, \urlprefix\url{https://doi.org/10.1007/s00340-005-1834-1}

\bibitem[{Everest and Atkinson(2008)}]{everest2008discrete}
Everest MA, Atkinson DB (2008) Discrete sums for the rapid determination of
  exponential decay constants. Review of Scientific Instruments 79(2):023108,
  \urlprefix\url{https://doi.org/10.1063/1.2839918}

\bibitem[{Bostrom et~al.(2015)Bostrom, Atkinson, and
  Rice}]{bostrom2015discrete}
Bostrom G, Atkinson D, Rice A (2015) The discrete fourier transform algorithm
  for determining decay constants—implementation using a field programmable
  gate array. Review of Scientific Instruments 86(4):043106,
  \urlprefix\url{https://doi.org/10.1063/1.4916709}

\bibitem[{Aboutanios(2009)}]{aboutanios2009estimation}
Aboutanios E (2009) Estimation of the frequency and decay factor of a decaying
  exponential in noise. IEEE Transactions on Signal Processing 58(2):501--509,
  \urlprefix\url{https://doi.org/10.1109/TSP.2009.2031299}

\bibitem[{Aboutanios(2011)}]{aboutanios2011estimating}
Aboutanios E (2011) Estimating the parameters of sinusoids and decaying
  sinusoids in noise. IEEE Instrumentation \& Measurement Magazine 14(2):8--14,
  \urlprefix\url{https://doi.org/10.1109/MIM.2011.5735249}

\bibitem[{Visschers et~al.(2021)Visschers, Wilson, Conneely, Mudrov, and
  Bougas}]{visschers2020rapid}
Visschers JC, Wilson E, Conneely T, Mudrov A, Bougas L (2021) Rapid parameter
  determination of discrete damped sinusoidal oscillations. Opt Express
  29(5):6863--6878, \urlprefix\url{https://doi.org/10.1364/OE.411972}

\bibitem[{van Veen \& S.~Leijnen(2019)}]{NNZoo}
van Veen \& S~Leijnen F (2019) The neural network zoo.
  \urlprefix\url{https://www.asimovinstitute.org/neural-network-zoo/}

\bibitem[{Bourlard and Kamp(1988)}]{bourlard1988auto}
Bourlard H, Kamp Y (1988) Auto-association by multilayer perceptrons and
  singular value decomposition. Biological cybernetics 59(4-5):291--294,
  \urlprefix\url{https://doi.org/10.1007/BF00332918}

\bibitem[{DeMers and Cottrell(1993)}]{demers1993non}
DeMers D, Cottrell GW (1993) Non-linear dimensionality reduction. In: Advances
  in neural information processing systems, pp 580--587

\bibitem[{Hinton and Salakhutdinov(2006)}]{hinton2006reducing}
Hinton GE, Salakhutdinov RR (2006) Reducing the dimensionality of data with
  neural networks. science 313(5786):504--507,
  \urlprefix\url{https://doi.org/10.1126/science.1127647}

\bibitem[{Theis et~al.(2017)Theis, Shi, Cunningham, and
  Husz{\'a}r}]{theis2017lossy}
Theis L, Shi W, Cunningham A, Husz{\'a}r F (2017) Lossy image compression with
  compressive autoencoders. arXiv preprint arXiv:170300395
  \urlprefix\url{https://arxiv.org/abs/1703.00395}

\bibitem[{Lu et~al.(2013)Lu, Tsao, Matsuda, and Hori}]{lu2013speech}
Lu X, Tsao Y, Matsuda S, Hori C (2013) Speech enhancement based on deep
  denoising autoencoder. In: Interspeech, vol 2013, pp 436--440

\bibitem[{Ng et~al.(2011)}]{ng2011sparse}
Ng A, et~al. (2011) Sparse autoencoder. CS294A Lecture notes 72(2011):1--19

\bibitem[{Asperti and Trentin(2020)}]{Asperti2020Balancing}
Asperti A, Trentin M (2020) Balancing reconstruction error and kullback-leibler
  divergence in variational autoencoders. IEEE Access 8:199440--199448,
  \doi{10.1109/ACCESS.2020.3034828}

\bibitem[{Gagliardi and Loock(2014)}]{gagliardi2014cavity}
Gagliardi G, Loock HP (2014) Cavity-enhanced spectroscopy and sensing, vol 179.
  Springer, \urlprefix\url{https://doi.org/10.1007/978-3-642-40003-2}

\bibitem[{Boens et~al.(2007)Boens, Qin, Basari{\'c}, Hofkens, Ameloot, Pouget,
  Lefevre, Valeur, Gratton, VandeVen et~al.}]{boens2007fluorescence}
Boens N, Qin W, Basari{\'c} N, Hofkens J, Ameloot M, Pouget J, Lefevre JP,
  Valeur B, Gratton E, VandeVen M, et~al. (2007) Fluorescence lifetime
  standards for time and frequency domain fluorescence spectroscopy. Analytical
  chemistry 79(5):2137--2149, \urlprefix\url{https://doi.org/10.1021/ac062160k}

\bibitem[{Cundall(2013)}]{cundall2013time}
Cundall R (2013) Time-resolved fluorescence spectroscopy in biochemistry and
  biology, vol~69. Springer Science \& Business Media

\bibitem[{Spiliotis et~al.(2020)Spiliotis, Xygkis, Klironomou, Kardamaki,
  Boulogiannis, Katsoprinakis, Sofikitis, and Rakitzis}]{spiliotis2020optical}
Spiliotis A, Xygkis M, Klironomou E, Kardamaki E, Boulogiannis G, Katsoprinakis
  G, Sofikitis D, Rakitzis T (2020) Optical activity of lysozyme in solution at
  532\,nm via signal-reversing cavity ring-down polarimetry. Chemical Physics
  Letters p 137345,
  \urlprefix\url{https://doi.org/10.1016/j.cplett.2020.137345}

\bibitem[{Papadakis et~al.(2011)Papadakis, Everest, Stamataki, Tzortzakis,
  Loppinet, and Rakitzis}]{papadakis2011development}
Papadakis V, Everest MA, Stamataki K, Tzortzakis S, Loppinet B, Rakitzis TP
  (2011) Development of cavity ring-down ellipsometry with spectral and
  submicrosecond time resolution. In: Instrumentation, Metrology, and Standards
  for Nanomanufacturing, Optics, and Semiconductors V, International Society
  for Optics and Photonics, vol 8105, p 81050L

\bibitem[{Yao and Pandit(1995)}]{yao1995cramer}
Yao YX, Pandit SM (1995) Cram{\'e}r-{R}ao lower bounds for a damped sinusoidal
  process. IEEE Transactions on signal processing 43(4):878--885,
  \urlprefix\url{https://doi.org/10.1109/78.376840}

\bibitem[{Chollet et~al.(2015)}]{chollet2015keras}
Chollet F, et~al. (2015) Keras. \url{https://keras.io}

\bibitem[{Wahl et~al.(2006)Wahl, Tan, Koulikov, Kharlamov, Rella, Crosson,
  Biswell, and Paldus}]{wahl2006ultra}
Wahl EH, Tan SM, Koulikov S, Kharlamov B, Rella CR, Crosson ER, Biswell D,
  Paldus BA (2006) Ultra-sensitive ethylene post-harvest monitor based on
  cavity ring-down spectroscopy. Optics express 14(4):1673--1684,
  \urlprefix\url{https://doi.org/10.1364/OE.14.001673}

\end{thebibliography}
\end{document}